\documentstyle[11pt,psfig]{article}

\begin{document}

\begin{center}
\LARGE{\bf Magnetic Susceptibility Study of the Heavy Rare Earth
Stannate Pyrochlores}
\end{center}

\bigskip
\begin{center}
V.~Bondah-Jagalu$^1$ and S.~T.~Bramwell$^2$
\end{center}

\bigskip

\noindent{$^1$ Department of Chemistry, The University of Edinburgh, The King's Buildings, West Mains Road, Edinburgh, EH9~3JJ, United Kingdom}\newline
\noindent{$^2$ Department of Chemistry, University College London, 20 Gordon Street, London, WC1H~0AJ, United Kingdom}

\bigskip

\noindent{Corresponding authors: e-mail: vin@ed.ac.uk and s.t.bramwell@ucl.ac.uk}

\bigskip

\begin{center}
\begin{minipage}{0.65\textwidth}
The series of magnetic rare earth pyrochlore stannates R$_2$Sn$_2$O$_7$
(R = rare earth, except Ce and Pm) have been investigated by powder susceptibility measurements down to
T $=1.8$ K. The results are compared to results for the analogous titanate series, which are well-known frustrated magnets.
Unlike the titanates, the whole series can be formed in the cubic pyrochlore structure. Possible experimental advantages of studying the stannates are discussed.
\end{minipage}
\end{center}

\bigskip

\noindent
\begin{minipage}{0.475\textwidth}
\section{Introduction}

The magnetic properties of the pyrochlore compounds can be very unusual on
account
of the high degree of frustration between antiferromagnetically coupled
magnetic ions
\cite{ander}. Cubic pyrochlore oxides A$_2$B$_2$O$_6$O' are described by the
face centred space group $Fd-3m$ where the trivalent A atoms are situated on
the 16d
sites, the tetravalent B atoms are on the 16c sites and the oxygen atoms are
in the
48f(O) and 8b(O') sites.  The A and the B atoms form identical
interpenetrating
sublattices that are spatially displaced from each other by $\langle
\frac{1}{2},\frac{1}{2},\frac{1}{2} \rangle$.  Each lattice can be described
as a
three dimensional array of corner linked tetrahedra and it is this basic
tetrahedral
motif that gives rise to the geometrical frustration of antiferromagnetic
interactions.
The A site is coordinated by two oxygen atoms (O') at the centre of the
tetrahedra and
by six oxygen atoms arranged in a puckered hexagonal ring around the metal
ions.
The structural details of pyrochlore compounds are well described in a very
comprehensive review of the oxide pyrochlores \cite{subra}. The tetrahedral
lattice is a feature of many other materials, examples including 

\end{minipage}
\hfill\
\begin{minipage}{0.475\textwidth}

ternary fluorides (e.g. NH$_4$CoAlF$_6$ \cite{raju}) and spinels
(e.g. MgCr$_2$O$_4$ \cite{shake}).\newline
\hspace*{0.4cm} Investigation of the magnetic properties of the pyrochlores has been
pioneered mainly
by J~.E.~Greedan and collaborators. Of particular interest is the discovery
\cite{ali} that
several compounds of the form R$_2$Mo$_2$O$_7$ (R = rare earth or Y)
show frozen in magnetic disorder in the apparent absence of chemical
disorder. For
these compounds the molybdenum(IV) ions are magnetic (d$^2$, S = $1$) and
the R
ion can be chosen to be nonmagnetic.  Depending on the rare earth, the
molybdates  may be  metallic  (R = Sm, Nd, Gd) or semiconducting (R = Y,
Tb-Yb) \cite{Greedan,taguc}
and it is not entirely clear to what extend local moment models of
frustration would apply to these materials. It therefore remains of
interest to characterise new families of magnetic pyrochlore materials with
localised
moments.  Here we describe a susceptibility study of a family of rare earth
stannates
R$_2$Sn$_2$O$_7$ in which the Sn(IV) ion is nonmagnetic and the rare-earth
ions
are chosen to be magnetic. This study follows on from an investigation of
the isostructural titanate materials R$_2$Ti$_2$O$_7$ \cite{bram1} that in turn
developed the early work of Cashion \cite{cashi}  and Bl\"{o}te \cite{blote}.

\end{minipage}
\twocolumn

While
investigating Ho$_2$Ti$_2$O$_7$ it was realised \cite{harri,bram2} that in
the
presence of local $\langle 1\,1\,1 \rangle$ easy axis anisotropy,
frustration arose from
a net {\em ferromagnetic} coupling between spins which has subsequently
been
established to be largely dipolar in origin \cite{Byron}.
The simple $\langle 1\,1\,1
\rangle$ Ising
ferromagnet was termed the ``spin ice'' model on account of its precise
mapping onto
the problem of proton disorder in ice. It gives a good first approximation
to the
properties of Ho$_2$Ti$_2$O$_7$ \cite{cashi} and Dy$_2$Ti$_2$O$_7$
\cite{ramir}. Following on from the titanates, we decided to look at the
magnetic
properties of the rare earth stannates, whose structural properties have
been well
studied \cite{whinf,briss1,kenne}. The compounds are cubic, with a linear
relationship between the radius of the lanthanide and the lattice. Non-cubic
pyrochlore
stannates do exist, such as Bi$_2$Sn$_2$O$_7$ \cite{roth,ravi}.

\section{Experimental}

Our samples were prepared from SnO$_2$ and R$_2$O$_3$ (R = La, Nd, Sm,
Eu, Gd, Dy, Ho, Er, Tm, Yb, Lu and Y) ordered from Strem and Aldrich, all
of a
minimum of 99.9\%
purity.  For the praseodymium and terbium stannate,
Pr$_6$O$_{11}$ and Tb$_4$O$_7$ were used respectively and it was assumed that
the
excess oxygen would escape as a gas during annealing.  It was not possible
to prepare
cerium stannate in this way, since the cerium remains as Ce$^{+4}$. However,
because of this fact, it would be possible to make La$_2$Ce$_2$O$_7$
\cite{bris2},
although this material adopts a disordered fluorite structure.  The oxides
were ground,
pressed into pellets and heated in air to $1400 ^o$C for 4 hours, this
length of time
being found sufficient for all of the compounds.  The X-ray diffraction
refinement of room temperature data gave
lattice constants consistent with
the literature values
\cite{kenne}.

Magnetisation was measured in the range $1.8$ K to $300$ K at an applied
field of
$10$ Oe ($T < 20$ K) and $100$ Oe (T $< 20$ K) using a
Quantum
Design SQUID magnetometer, and divided by field and amount of substance to
give
the susceptibility per mole of lanthanide. The low field susceptibility was
scaled to fit
the higher field data, as the field positioning at $10$ Oe is not very
accurate. Zero-
field cooled versus field cooled susceptibility was determined in an applied
field of
$10$ Oe between T = $1.84$ K and T = $20$ K, and magnetisation versus field
isotherms were measured at T = $1.8$ K in applied fields of up to $7$ T.
Estimated
Curie-Weiss temperatures and magnetic moments
are listed
in table 1.
Individual curves of effective moment $\mu_{eff} = \sqrt{8\chi T}$ versus
temperature are shown in figure 1.
\begin{figure}[h]
\centerline{
\psfig{figure=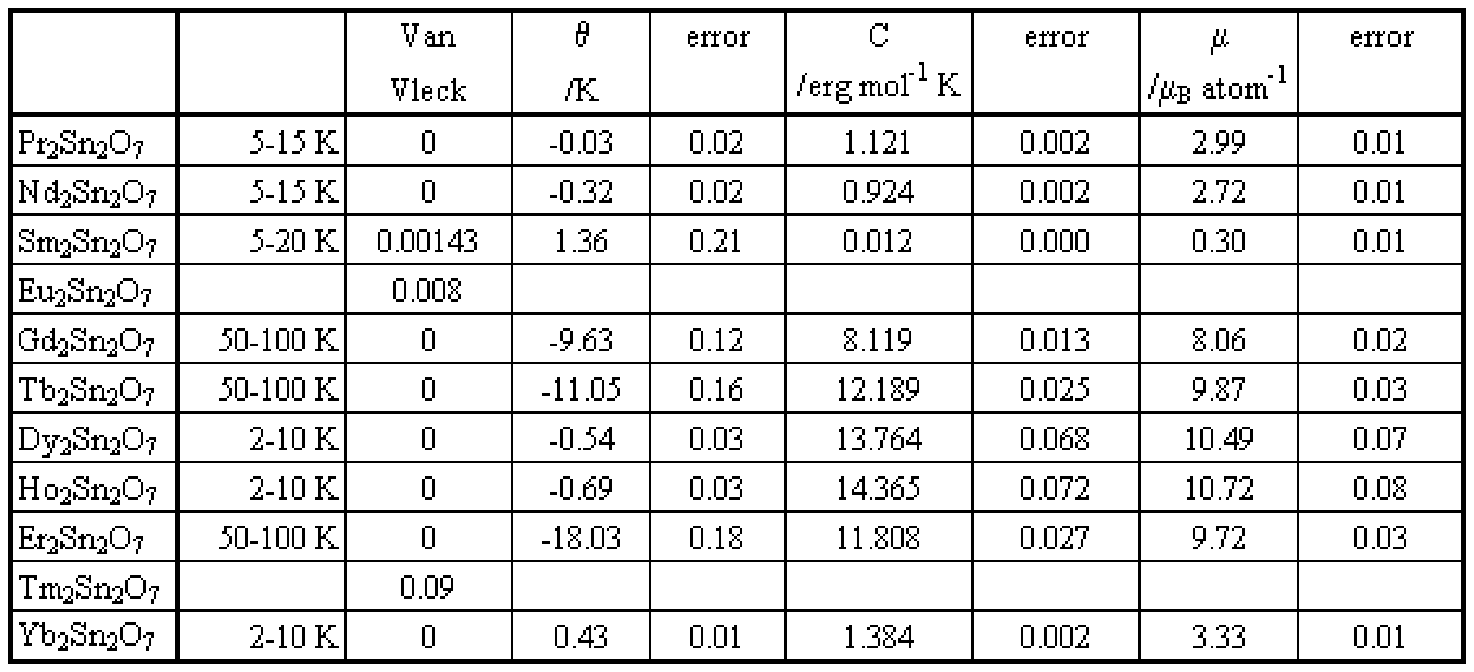,height=5.4cm,width=8.5cm,angle=0}}
\small {Table 1 : The Curie-Weiss temperature, constant and magnetic moment for 
the rare earth pyrochlore stannates, estimated by extrapolating data from the given 
temperature ranges and ``if required'' applying the Van Vleck paramagnetism correction.}
\end{figure}

\begin{figure}[h]
\centerline{
\psfig{figure=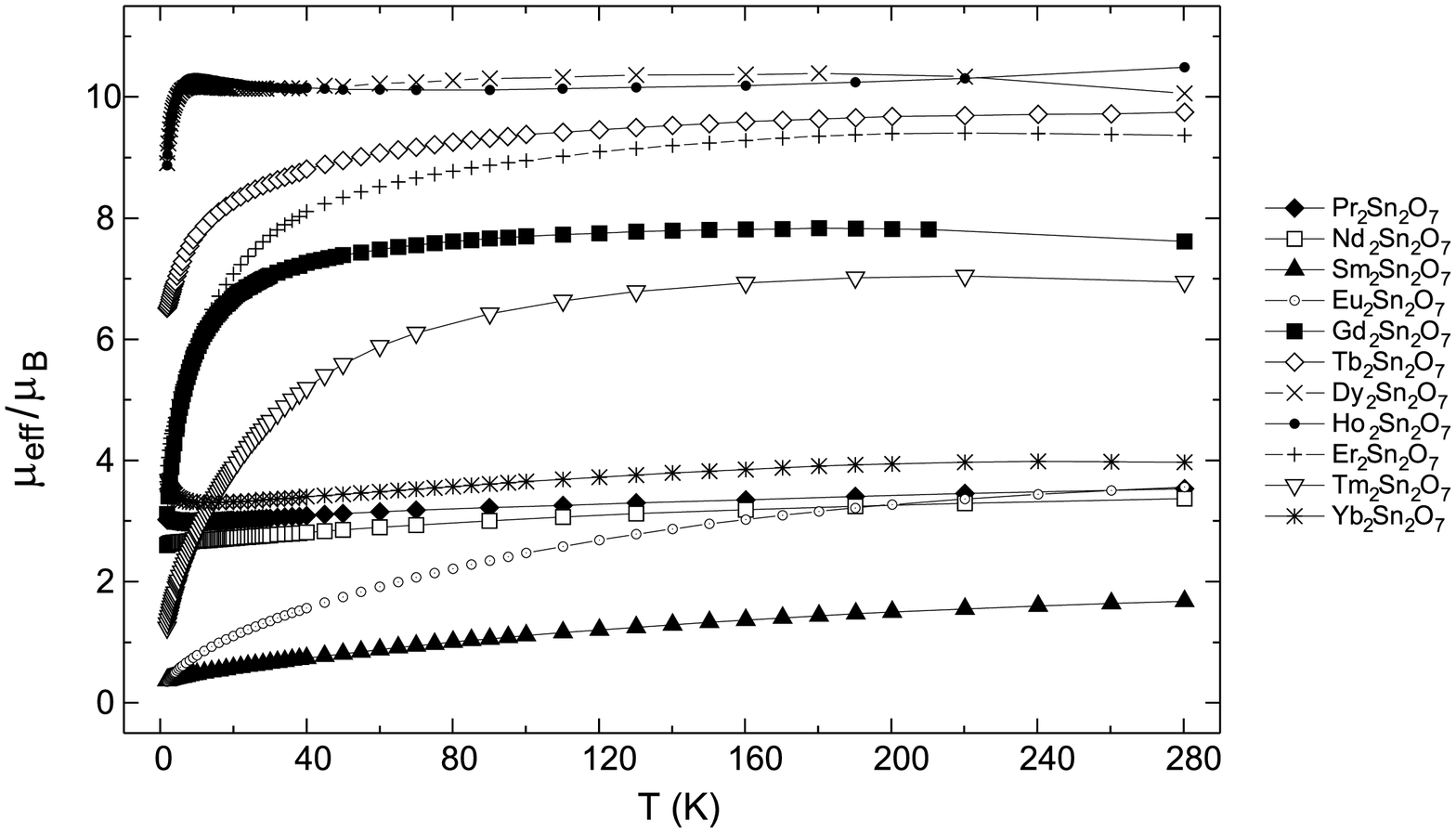,height=6.2cm,width=8.6cm,angle=0}}
\small {Figure 1 : The effective moments $\mu_{eff}$ versus temperature $T$ for the rare
earth pyrochlore stannates. The moments are derived from measured molar
susceptibility
$\chi$  (in cgs units) via the relation $\mu_{eff} = \sqrt{8 \chi T}$.}
\end{figure} 

\section{Results and Discussion}

$R_2Sn_2O_7$ (R = Nd, Tb, Dy, Ho, Er, Yb) almost certainly have
doublet ground states with strong single ion anisotropy. The latter is
evidenced by the
fact that the magnetisation approaches saturation at roughly half the free
ion value.
Amongst this group all except Yb$_2$Sn$_2$O$_7$ have Curie constants that
are
consistent with a ground state for $R^{3+}$ consisting mainly of M$_J$ =
$\pm$ J
for the free ion ground term.  For R = Nd, Gd, Dy, Ho, Er, Tm, Yb
the
magnetic behaviour is extremely close to that of the analogous titanate.
Thus one
expects ``dipolar spin ice'' behaviour for R = Ho, Dy and ordering at
T $<
1.8$ K for R = Nd, Er, Yb, while R = Tm clearly has a singlet
ground
state derived from crystal field splitting of the free ion term
\cite{Byron,blote,zinki}. Both
Er$_2$Sn$_2$O$_7$ and Er$_2$Ti$_2$O$_7$ exhibit a pronounced FC-ZFC
splitting in the susceptibility at about $3.4$ K. This feature needs to be
investigated
further. For Tb$_2$Sn$_2$O$_7$, the Curie-Weiss temperature $\theta_{CW}$  =
$-
11.1$ K, is apparently much less than in Tb$_2$Ti$_2$O$_7$, $\theta_{CW}$  =
$-
19$ K \cite{gard1}.  It is possible that our sample of Tb$_2$Sn$_2$O$_7$ is
slightly
non-stoichiometric. Consistent with this, we observe a slight FC-ZFC
splitting in the
susceptibility below $4.5$ K, which is not reported for the titanate.
The latter has been described as a
co-operative
paramagnet with no sign of magnetic order down to T = $0.07$ K \cite{gard2}.

There have not been detailed reports of the magnetism of titanates analogous
to
R$_2$Sn$_2$O$_7$ with R = Pr, Sm, Eu, so we consider their behaviour
in
more detail.

$Pr_2Sn_2O_7$: The saturation magnetisation and effective moment
approach the free ion value, while the Curie-Weiss constant is small. This
suggests
that crystal field effects and magnetic coupling are small in this material.
This
behaviour can be contrasted with Pr$_2$O$_3$
\cite{kern,vicke} which has a
large
$\theta_{CW}$ $\approx$ -60 K arising from the crystal field splitting of the
H$_6$
free ion term of Pr$^{3+}$ to give a ground state singlet and low lying
doublet.

$Sm_2Sn_2O_7$: The effective moment versus temperature appears
to
have a term in (temperature)$^{\frac{1}{2}}$ dominating at high temperature,
suggestive of a Van Vleck paramagnetism. It is typical for the
susceptibility of
Sm$^{3+}$ salts to be dominated by the Van Vleck term near room temperature,
and
to show increasingly Curie-Weiss behaviour at low temperature \cite{borov}.
Ascribing the value $0.00143$ erg Oe$^{-2}$ mol$^{-1}$ to the Van Vleck term
gives a Curie-Weiss temperature of $1.36$ K and moment derived from the
Curie
constant of $0.30$, which is less than the moment of the
$^6$H$_{\frac{5}{2}}$ free
ion term,  $0.84~\mu_B$ atom$^{-1}$. This reduction in moment is consistent
with the
ground term in Sm$_2$Sn$_2$O$_7$ being a doublet dominated by free ion M$_J$
levels with M$_J$ significantly less than the maximum, $\frac{5}{2}$. A
similar
effect is observed in Sm$_2$O$_3$, which shows a similar susceptibility
curve
\cite{borov}.

$\it {Eu_{2}Sn_{2}O_{7}}$:   The susceptibility is roughly constant below
100 K,
behaviour typical for Eu$^{3+}$, in which the free ion $^7$F$_0$ ground term
is
non-magnetic, leaving only the Van Vleck term. Similar behaviour is observed
in
Eu$_2$O$_3$ \cite{borov}. The observed magnitude of the Van Vleck term
($\sim0.008$ erg Oe$^{-2}$ mol$^{-1}$) in Eu$_2$Sn$_2$O$_7$ is close to that
of
Eu$_2$O$_3$, which has been shown to be enhanced by anisotropic exchange
between the Eu ions \cite{huang}.

\section{Concluding Remarks}

The rare earth stannates can be prepared in the cubic form across the whole
lanthanide
series and make an ideal system for the comparative study of magnetic
properties.
Experimental investigation of the series to lower temperatures is in
progress. The
presence of Sn in the structure will facilitate experimental investigation
of these
materials by Mossbauer $^{119}$Sn spectroscopy.  For example, using this
technique,
Sn$_2$B$_2$O$_7$ (B = Nb, Ta) have been shown to be actually of the
form
Sn$_{2-2u}^{2+}$(Sn$_{2u}^{4+}$B$_{2-u}^{5+}$)O$_{7-(5/2)u}$
; $\it {u}$ = $0.18-0.20$) \cite{stewa}. Solid solutions of the type
R$_x$R'$_{1-x}$Sn$_2$O$_7$ (R = rare earth, R' = Ln, Y) can
also
be formed. In this context, $^{119}$Sn MAS NMR spectroscopy has been shown to
be
a particularly powerful probe of local cation geometry on account of the
very large
$^{119}$Sn chemical shifts \cite{grey}.  We finally note that, to complete the
series,
cerium stannate pyrochlore (Ce$_2$Sn$_2$O$_7$) has recently been prepared
for the
first time \cite{tolla}.

\end{document}